\documentclass[12pt]{article}

\usepackage{amssymb,amsmath} 
\usepackage{bm}
\usepackage{epsfig} 

\newcommand{\beqs}{\begin{equation*}}
\newcommand{\beq}{\begin{equation}}

\newcommand{\eeqs}{\end{equation*}}
\newcommand{\eeq}{\end{equation}}

\newcommand{\beqas}{\begin{eqnarray*}}
\newcommand{\beqa}{\begin{eqnarray}}

\newcommand{\eeqas}{\end{eqnarray*}}
\newcommand{\eeqa}{\end{eqnarray}}

\newcommand{\eq}[2]{\begin{equation} #1 \label{#2} \end{equation}}

\newcommand{\la}{\lambda}
\newcommand{\si}{\sigma}

\newcommand{\blist}{\begin{itemize}}

\newcommand{\elist}{\end{itemize}}

\providecommand{\href}[2]{#2}

\DeclareFontFamily{OT1}{rsfs}{}
\DeclareFontShape{OT1}{rsfs}{m}{n}{ <-7> rsfs5 <7-10> rsfs7 <10->rsfs10}{} 
\DeclareMathAlphabet{\mycal}{OT1}{rsfs}{m}{n}

\DeclareMathOperator{\extdm}{d}
\newcommand{\extd}{\extdm \!}

\newcommand{\KK}{k}
\newcommand{\pont}{\mathcal{P}}

\begin{document}





 






\begin{titlepage}

{\hfill MIT-CTP 3901}

\begin{center}

\vspace*{1.5truecm}

\textbf{\Large Einstein-Weyl from Kaluza-Klein}

\vspace{10ex}

D.~Grumiller\footnote{e-mail: {\tt grumil@lns.mit.edu}} and R.~Jackiw\footnote{e-mail: {\tt jackiw@lns.mit.edu}}

  \vspace{7ex}

{\em Center for Theoretical Physics,
Massachusetts Institute of Technology,\\
77 Massachusetts Ave.,
Cambridge, MA  02139}

\end{center}

\vspace{14ex}

\begin{abstract}

We discuss the Kaluza-Klein reduction of spaces with (anti-)self-dual Weyl tensor and point out the emergence of the Einstein-Weyl equations for the reduction from four to three dimensions. As a byproduct we get a simple expression for the gravitational instanton density in terms of the Kaluza-Klein functions.

\end{abstract}

\end{titlepage}

\setcounter{footnote}{0}


\section{Introduction}

Recently we carried out a Kaluza-Klein reduction from $n$ to $n-1$ dimensions of conformal tensors (Weyl for $n\geq 4$, Cotton for $n\geq 3$) \cite{Grumiller:2006ww}. We obtained the descendant expressions in terms of the Kaluza-Klein functions (metric tensor and gauge potential in the lower dimensionality). Further we imposed the condition of conformal flatness, i.e., the vanishing of the higher dimensional conformal tensor, thereby obtaining equations satisfied by the Kaluza-Klein functions. Solutions to these equations describe the immersion of a lower dimensional structure into a conformally flat space.

When reporting our calculations at a conference \cite{Jackiw:2007ge}, we were apprised that our $4\to 3$ dimensional story is closely related to the theory of Einstein-Weyl spaces in three dimensions, widely studied in mathematics, though apparently of no relevance to physics \cite{Bergmann:1942}. We were informed that our final equations and results are known to mathematicians \cite{Jones:1985,Tod:1992,LeBrun:1999,Pederson:1993,Eastwood:1997,Calderbank:2000tk}, 
provided some adjustments are made. (We studied spaces with Lorentzian signature, which is not common practice in the mathematical setting.) Nevertheless, it appears that our analysis, if not our results, is somewhat different from what is found in the mathematical literature. Also the interest in Einstein-Weyl theory is mostly non-existent in physics. Therefore, in this paper we describe the material, with the hope that it will appeal both to physicists and to mathematicians.

In Section \ref{se:2} the Einstein-Weyl theory is reviewed. In Section \ref{se:3}, dimensional reduction of the 4-dimensional Weyl tensor is accomplished by the Kaluza-Klein method. Self-duality conditions in Euclidean four dimensions then lead to equations that are closely related to the Einstein-Weyl equations in three dimensions. Conformal flatness is then reconsidered as a more restrictive condition. With Lorentzian signature, conformal self-duality is not possible with real fields; only conformal flatness can be imposed. We exhibit the differences that arise when Lorentzian signature is employed. In Section \ref{se:4} we present a simple result that follows from our Kaluza-Klein reduction of the Weyl tensor and its dual: the gravitational instanton density (also known as Chern-Pontryagin term) is expressed in terms of the Kaluza-Klein functions. Finally, we discuss an application thereof in physics, to Chern-Simons modified gravity.

\section{Pr{\'e}cis of Einstein-Weyl Theory}\label{se:2}

\newcommand{\WD}{\nabla}

Einstein-Weyl theory (in any dimension) is equipped with a metric tensor $g_{\mu\nu}$ and an additional vector $w_\mu$ -- the ``Weyl potential'' -- which arises when the covariant ``Weyl derivative'' $\WD^W_\mu$, involving the torsion-less ``Weyl connection'' $w^\lambda{}_{\mu\nu}$, acts on $g_{\mu\nu}$ and preserves its conformal class, cf.~e.g.~\cite{LeBrun:1999}.
 \eq{
\WD^W_\lambda g_{\mu\nu} := \partial_\lambda g_{\mu\nu} - w^\sigma{}_{\lambda\mu}g_{\sigma\nu} - w^\sigma{}_{\lambda\nu}g_{\mu\sigma} = 2w_\lambda g_{\mu\nu}
}{eq:EW1}
The Weyl connection, which leads to \eqref{eq:EW1}, can be constructed from the conventional Christoffel connection $\Gamma^\lambda{}_{\mu\nu}$, supplemented by a $w_\mu$-dependent expression.
\eq{
w^\lambda{}_{\mu\nu}:=\Gamma^\lambda{}_{\mu\nu}+w^\lambda g_{\mu\nu}-w_\mu\delta^\lambda_\nu-w_\nu\delta^\lambda_\mu
}{eq:EW2}

A curvature tensor is determined as usual by
\eq{
\left[\WD_\mu^W,\WD_\nu^W\right]V_\alpha = - {^W}\!r^\beta{}_{\alpha\mu\nu}V_\beta
}{eq:EW3}
whose traces define ``Ricci'' quantities.
\begin{align}
{^W}\!r_{\mu\nu} & := {^W}\!r^\alpha{}_{\mu\alpha\nu}\\
{^W}\!r & := {^W}\!r^\mu{}_\mu
\end{align}

The Einstein-Weyl equation then requires that ${^W}\!r_{(\mu\nu)}$, the symmetric part of the ``Ricci'' tensor,\footnote{We define symmetrization by $r_{(\mu\nu)}:=\frac12 (r_{\mu\nu}+r_{\nu\mu})$, anti-symmetrization by $r_{[\mu\nu]}:=\frac12 (r_{\mu\nu}-r_{\nu\mu})$ and note that generically ${^W}\!r_{\mu\nu}$ is not symmetric.} be in the same conformal class as the metric tensor,
\eq{
{^W}\!r_{(\mu\nu)} = \lambda g_{\mu\nu}
}{eq:EW4}
or equivalently in three dimensions
\eq{
{^W}\!r_{(\mu\nu)}-\frac13 g_{\mu\nu} {^W}\!r = 0\,.
}{eq:EW5}
From \eqref{eq:EW2} and \eqref{eq:EW3} ${^W}\!r_{(\mu\nu)}$ can be expressed in terms of the usual Ricci tensor $r_{\mu\nu}$, supplemented by $w_\mu$-dependent terms.
\eq{
{^W}\!r_{(\mu\nu)} = r_{\mu\nu} + d_{(\mu} w_{\nu)} + w_\mu w_\nu + g_{\mu\nu} (d_\lambda w^\lambda - w_\lambda w^\lambda)
}{eq:EW6}
Here $d$ is the covariant derivative constructed with the 3-dimensional Christoffel connection. Thus the Einstein-Weyl equation \eqref{eq:EW5} requires the vanishing of a tracefree quantity.
\eq{
r_{\mu\nu}-\frac13 g_{\mu\nu} r + d_{(\mu}w_{\nu)}-\frac13 g_{\mu\nu} d_\la w^\la + w_\mu w_\nu - \frac13 g_{\mu\nu} w_\la w^\la = 0
}{eq:EW7}

The equations \eqref{eq:EW1} and \eqref{eq:EW6} are preserved under conformal transformations: the metric tensor is rescaled and the Weyl potential undergoes a gauge transformation.
\eq{
g_{\mu\nu}\to e^{2\sigma}g_{\mu\nu}\,,\qquad w_\mu\to w_\mu + \partial_\mu\sigma
}{eq:EW8}
This gauge freedom is fixed by choosing the ``Gauduchon gauge'' $d_\mu w^\mu=0$. Within the Gauduchon gauge, a further calculation shows that \eqref{eq:EW7} may be simplified. First present \eqref{eq:EW7} as 
\eq{
r_{\mu\nu}+d_{(\mu}w_{\nu)}+w_\mu w_\nu = \Lambda g_{\mu\nu}\,.
}{eq:EW9}
Multiply this by $d^\mu w^\nu$ to form
\begin{subequations}
\begin{multline}
d^{(\mu}w^{\nu)}d_{(\mu}w_{\nu)} = \Lambda d_\mu w^\mu - r_{\mu\nu}d^\mu w^\nu - w_\mu w_\nu d^\mu w^\nu \\
= \Lambda d_\mu w^\mu - d^\mu(r_{\mu\nu}w^\nu) - \frac12 w^\mu d_\mu w^2 + w^\nu d^\mu r_{\mu\nu}\,. 
\label{eq:EW10}
\end{multline}
Since $d^\mu r_{\mu\nu}=\frac12 \partial_\nu r$, the above is rewritten as
\eq{
d^{(\mu}w^{\nu)}d_{(\mu}w_{\nu)} = \big(\Lambda - \frac12 r + \frac12 w^2\big)d_\mu w^\mu - d^\mu \big(r_{\mu\nu}w^\nu-\frac12 w_\mu r-\frac12 w_\mu w^2\big)\,.
}{eq:EW11}
\end{subequations}
In the Gauduchon gauge, the first term on the right hand side vanishes. The second term vanishes when integrated over the relevant manifold, provided it is compact. Alternatively, if the manifold is open, with a boundary at infinity, sufficiently rapid drop-off conditions on the relevant quantities still ensure a vanishing integral. In either case, the integral of the left hand side vanishes. If the metric on the space is positive, the vanishing of the integral ensures the vanishing of the integrand and finally of $d_{(\mu}w_{\nu)}$. In this situation the Einstein-Weyl equations, gauge-fixed to the Gauduchon gauge, reduce to
\eq{
r_{\mu\nu}-\frac13 g_{\mu\nu}r + w_\mu w_\nu - \frac13 g_{\mu\nu} w_\lambda w^\lambda = 0\,,
}{eq:EW12} 
\eq{
d_{(\mu}w_{\nu)}=0\,.
}{eq:EW13}
Equation \eqref{eq:EW13} shows that in the Gauduchon gauge the Weyl vector $w_\mu$ is a Killing vector for the 3-dimensional Einstein-Weyl geometry, with the above delineated further properties of the 3-space.

\section{Kaluza-Klein Reduction of the 4-dimensional Weyl Tensor and its Dual}\label{se:3}

We are concerned with the 4-dimensional Weyl tensor, 
\eq{
C^{KLMN}:=R^{KLMN}-g^{K[M}S^{N]L}+g^{L[M}S^{N]K}
}{eq:EW16}
which is constructed from the Riemann tensor
\eq{
R^K{}_{LMN} := \partial_M \Gamma^K{}_{NL} - \partial_N \Gamma^K{}_{ML} + \Gamma^K{}_{MP}\Gamma^P{}_{NL}-\Gamma^K{}_{NP}\Gamma^P{}_{ML}
}{eq:EW14}
and the Schouten tensor 
\eq{
S_{MN} := R_{MN}-\frac16 g_{MN} R\,,
}{eq:EW15}
where
\eq{
R_{MN} := R^K{}_{MKN}\,,\qquad R:= g^{MN}R_{MN}\,.
}{eq:EW17}
We use capital letters to denote 4-dimensional quantities, as above, and lower case letters for 3-dimensional entities, as in Section \ref{se:2}.

We choose the 4-dimensional metric tensor $g_{MN}$ to be of the Kaluza-Klein form
\begin{equation}
  \label{eq:j1}
  g_{MN}=e^{2\si}\left(\begin{array}{cc}
  g_{\mu\nu}+a_\mu a_\nu & a_\mu \\
  a_\nu & 1  
   \end{array}\right)\,.
\end{equation}
corresponding to the line element
\begin{equation}
  \label{eq:le2}
  \extd s^2_{(4)}=g_{MN}\extd x^M\extd x^N=e^{2\sigma}\left[\extd s^2_{(3)}+(a_\mu\extd x^\mu+\extd x^4)^2\right]\,,
\end{equation}
with
\eq{
\extd s^2_{(3)}=g_{\mu\nu}dx^\mu dx^\nu\,.
}{eq:EW18}
Since we are interested in the conformal tensor the overall conformal factor $e^{2\sigma}$ has no significant role, so henceforth we omit it. Furthermore we take the Kaluza-Klein mode functions ($g_{\mu\nu}$, $a_\mu$) to be independent of the ``fourth'' coordinate denoted by $x^4$.

The Riemann tensor, evaluated on the metric \eqref{eq:j1} is given by a variant of the Gauss-Codazzi equations. These then lead to the corresponding formulas for the Weyl tensor
\begin{align}
\label{eq:W1}
& C^{\mu\nu\lambda\tau} =2\left(g^{\mu[\lambda}c^{\tau]\nu}-g^{\nu[\lambda}c^{\tau]\mu}\right)= -\epsilon^{\mu\nu\alpha}\epsilon^{\lambda\tau\beta}c_{\alpha\beta} \\
\label{eq:W2}
& c^{\mu\nu} := \frac12 \left(r^{\mu\nu}-\frac13 g^{\mu\nu}r+f^\mu f^\nu-\frac13 g^{\mu\nu}f^2\right) \\
\label{eq:W3}
& C^{\mu\nu\lambda 4} + C^{\mu\nu\lambda\tau}a_\tau =-\epsilon^{\mu\nu\tau}{\KK}^{\lambda}{}_\tau \\
\label{eq:W4}
& {\KK}_{\lambda\tau} := d_{(\lambda} f_{\tau)}
\end{align}
with
\eq{
f^\lambda := \epsilon^{\lambda\mu\nu}\partial_\mu a_\nu\,.
}{eq:EW19}
The quantity $\epsilon^{\mu\nu\tau}$ denotes the $\epsilon$-tensor, which is related to the antisymmetric $\epsilon$-symbol $\tilde{\epsilon}^{\mu\nu\tau}$ by $\epsilon^{\mu\nu\tau}=\tilde{\epsilon}^{\mu\nu\tau}/\sqrt{g}$, and $d_\mu$ is the 3-dimensional covariant derivative involving the 3-dimensional Christoffel connection. Note that both $c_{\mu\nu}$ and ${\KK}_{\mu\nu}$ are traceless. 

Now we define the dual Weyl tensor ($\epsilon^{MNRS}$ again is the tensor).
\eq{
{{^\ast}\! C}^{ABMN} := \frac{1}{2}\epsilon^{MNRS} C^{AB}{}_{RS}
}{eq:EW20}
The Weyl tensor and its dual share all the symmetries of the Riemann tensor. Also they are traceless in every pair of indices. Moreover, in four dimensions not only $C^A{}_{BCD}$ is conformally invariant and thus independent from $\sigma$ in \eqref{eq:le2}, but also its dual (with the same index positions).

The relations between the 3-dimensional components of ${^\ast}\! C^{ABMN}$ and $C^{ABMN}$ are 
\begin{align}
& {^\ast}\! C^{\sigma\tau\mu\nu} = \epsilon^{\mu\nu\alpha} g_{\alpha\beta} \left(C^{\sigma\tau\beta 4}+C^{\sigma\tau\beta\lambda}a_\lambda\right) 
\label{eq:dualW1} \\
& {^\ast}\! C^{\sigma\tau\mu 4} + {^\ast}\! C^{\sigma\tau\mu\nu}a_\nu = \frac12 \epsilon^{\mu\alpha\beta}g_{\alpha\gamma}g_{\beta\delta} C^{\sigma\tau\gamma\delta} \,.
\label{eq:dualW2}
\end{align}
The remaining components of ${^\ast}\! C^{ABMN}$ are determined by the symmetries and trace properties of that tensor.

We now equate \eqref{eq:dualW1} and \eqref{eq:dualW2} to ($\pm$) the corresponding Weyl tensor components thereby requiring the 4-dimensional Weyl tensor be (anti-)self-dual. This produces equations that are solved by
\eq{
c_{\mu\nu} = \pm {\KK}_{\mu\nu}\,.
}{eq:mainResult}

Comparison with the Einstein-Weyl equations \eqref{eq:EW7} shows that we have regained them, provided $f^\mu$ is identified with $\pm w^\mu$.  Moreover, we are already in the Gauduchon gauge, by virtue of the transversality of $f^\mu$, see \eqref{eq:EW19}. We may appeal to asymptotic conditions to argue that ${\KK}_{\mu\nu}$ vanishes, as above. Alternatively, the demand that the 4-dimensional space be conformally flat, i.e., that its Weyl tensor vanishes so that it is both self-dual and anti-self-dual, implies that $c_{\mu\nu}$ and ${\KK}_{\mu\nu}$ vanish separately. Therefore, the asymptotic conditions which establish ${\KK}_{\mu\nu}=0$ are strong enough to render conformally flat any {(anti-)}\-self-dual spacetime with a Killing vector [given by $\partial_{x^4}$ in the adapted coordinate system \eqref{eq:le2}].

Once \eqref{eq:mainResult} is replaced by the vanishing of each side, it is a straightforward matter to derive further equations that also appear in the mathematics literature \cite{Jones:1985,Tod:1992,Pederson:1993,Eastwood:1997,Calderbank:2000tk}
\eq{
r = 5f^2 + c\,,
}{eq:EW21}
\eq{
d_{(\mu}F_{\nu)}=0\,,
}{eq:EW22}
where $c$ is a constant and
\eq{
F^\mu := \epsilon^{\mu\nu\lambda}d_\nu f_\lambda\,.
}{eq:EW23}
Equation \eqref{eq:EW22} shows that there exists in the 3-dimensional geometry a further Killing vector, $F^\mu$, which is constructed from the curl of $f_\mu$, when the latter is non-vanishing.

\enlargethispage{1cm}

When the spacetime possesses Lorentzian signature, the Gauduchon argument cannot be carried to the conclusion that $d_{(\mu} w_{\nu)}$ vanishes. However our dimensional reduction procedure arrives at that result directly. With Lorentzian signature \eqref{eq:j1} is replaced by
\begin{equation}
  \label{eq:j1old}
  g_{MN}=e^{2\si}\left(\begin{array}{cc}
  g_{\mu\nu}-a_\mu a_\nu & -a_\mu \\
  -a_\nu & -1  
   \end{array}\right)\,.
\end{equation}
Formulas \eqref{eq:W1}, \eqref{eq:W3}, \eqref{eq:W4} and \eqref{eq:EW19} continue to hold but \eqref{eq:W2}  changes in that the terms quadratic in $f^\mu$ acquire the opposite sign. With Lorentzian signature (anti-)self-duality cannot be imposed on real fields, so the only possible requirement is vanishing of the ($3+1$)-dimensional Weyl tensor. This leads to the vanishing of $c_{\mu\nu}$ (with the appropriate sign change) and to the Killing equation for $f^\mu$.

Finally we observe that it is not known whether the Einstein-Weyl equations derive from an action/Lagrangian. Our approach does not shed any new light on this. However, when a further Ansatz is posited for our equations, viz.~that the Kaluza-Klein functions be circularly symmetric, 2-dimensional actions that lead to these equations have been constructed \cite{Grumiller:2006ww}. These actions are related to each other by a specific duality that exists for generic 2-dimensional dilaton gravity \cite{Grumiller:2006xz}.

\section{Chern-Pontryagin Term}\label{se:4}

The Chern-Pontryagin term
\eq{
\pont:=\frac12\, {^\ast}\! R^{ABCD}\,R_{ABCD}\,, \qquad {^\ast}\! R^{ABCD}:=\frac12 \epsilon^{CDMN}R^{AB}{}_{MN}\,.
}{eq:EW37}
can be represented by the alternative formula
\eq{
\pont=\frac12\, {^\ast}\! C^{ABCD}\,C_{ABCD}\,.
}{eq:EW38}
Its properly normalized volume integral yields the gravitational instanton number. 

With the results from Section \ref{se:3} it is now straightforward to calculate the dimensional reduction of $\pont$. Using the Kaluza-Klein split \eqref{eq:j1} [or \eqref{eq:j1old}] it proliferates into
\eq{
\pont = \frac12\,{^\ast}\! C^{\alpha\beta\gamma\delta}\,C_{\alpha\beta\gamma\delta} + 2\, {^\ast}\! C^{\alpha\beta\gamma 4}\,C_{\alpha\beta\gamma 4} + 2\, {^\ast}\! C^{\alpha 4 \beta 4}\,C_{\alpha 4 \beta 4}\,.
}{eq:EW24}
By virtue of the symmetry- and trace-properties of $C_{ABCD}$ and its dual, we obtain from \eqref{eq:W1}-\eqref{eq:EW19}, \eqref{eq:dualW1} and \eqref{eq:dualW2} the simple result
\eq{
\pont = 8 c^{\mu\nu}{\KK}_{\mu\nu}
}{eq:EW25}
for the Chern-Pontryagin term. This formula is useful for Chern-Simons modified gravity \cite{Jackiw:2003pm}. Namely, in that theory $\pont$ has to vanish on classical solutions. In practice it turns out to be difficult to implement this constraint effectively \cite{ariadna}. However, if 4-dimensional space-time admits one Killing vector our reduction scheme applies and \eqref{eq:EW25} can be exploited. 

The constraint
\eq{
\pont = 0 = c^{\mu\nu}{\KK}_{\mu\nu}
}{eq:EW26}
has three different classes of solutions. Either $c_{\mu\nu}$ vanishes or ${\KK}_{\mu\nu}$ vanishes or they are orthogonal, in the sense that \eqref{eq:EW26} holds. This parallels the situation in gauge theory, where ${^\ast}\!F F \propto\bm{E \cdot B}$ vanishes either for electric ($\bm{E}\neq 0$, $\bm{B}=0$), magnetic ($\bm{B}\neq 0$, $\bm{E}=0$) or wave configurations ($\bm{E \cdot B}=0$, $\bm{E}\neq 0 \neq \bm{B}$).\footnote{The analogy with gauge theory also applies to the square of the Weyl tensor and its dual, 
$C^{ABCD}\, C_{ABCD} = {^\ast}\!C^{ABCD}\, {^\ast}\!C_{ABCD} = 8 \left(c^{\mu\nu}c_{\mu\nu}\pm{\KK}^{\mu\nu}{\KK}_{\mu\nu}\right)$, 
which matches with the gauge theoretic $F^2\propto (\bm{E}^2\pm\bm{B}^2)$, where the upper (lower) sign refers to Euclidean (Lorentzian) signature.} 

The ``electric'' case, $c_{\mu\nu}\neq 0$ and ${\KK}_{\mu\nu}=0$ is equivalent to the Killing equation
\eq{
d_{(\mu}f_{\nu)}=0\,,
}{eq:EW28}
which means that for non-vanishing $f_\mu$ the 4-dimensional space must exhibit at least two Killing vectors: one of them, $\partial_{x^4}$, is assumed for the Kaluza-Klein reduction, while the other emerges from lifting $f^\mu$ to a 4-dimensional Killing vector. However, with non-vanishing $c_{\mu\nu}$ the vector $F^\mu$ from \eqref{eq:EW23} in general does not fulfill the Killing equation \eqref{eq:EW22}. If $f^\mu$ is geodesic, $f^\nu d_\nu f^\mu=0$, then the Killing equation \eqref{eq:EW28} establishes a conservation equation
\eq{
d_\mu \,j = 0
}{eq:EW35}
for the scalar current 
\eq{
j = f^2\,.
}{eq:EW36}
This conservation is neither necessary nor sufficient for \eqref{eq:EW28}.

The ``magnetic'' case, ${\KK}_{\mu\nu}\neq 0$ and $c_{\mu\nu}=0$, yields a condition resembling the Einstein equations,
\eq{
r^{\mu\nu}-\frac13 g^{\mu\nu}r\pm f^\mu f^\nu\mp\frac13 g^{\mu\nu}f^2 = 0\,.
}{eq:EW27}
The upper (lower) sign is valid for Euclidean (Lorentzian) signature. Instead of \eqref{eq:EW21}, which no longer needs to hold, the Bianchi identities establish a covariant conservation equation 
\eq{
d_\mu \,j^{\mu\nu} = 0
}{eq:EW33}
for the symmetric tensor current
\eq{
j^{\mu\nu} = g^{\mu\nu}(r\mp 2f^2)\pm6f^\mu f^\nu\,.
}{eq:EW34}
This conservation is necessary but not sufficient for \eqref{eq:EW27}.

The general case, $c^{\mu\nu}{\KK}_{\mu\nu}=0$ and $c_{\mu\nu}\neq 0 \neq {\KK}_{\mu\nu}$, allows further analysis. Inserting \eqref{eq:W2} and \eqref{eq:W4} into \eqref{eq:EW26} yields
\eq{
\left(r^{\mu\nu}-\frac13 g^{\mu\nu}r \pm f^\mu f^\nu\mp\frac13 g^{\mu\nu}f^2\right)d_\mu f_\nu = 0\,.
}{eq:EW29}
Again the upper (lower) sign is valid for Euclidean (Lorentzian) signature.
Now we use $d_\mu f^\mu=0$ and get
\eq{
r^{\mu\nu}d_\mu f_\nu \pm \frac12 d_\mu \left(f^2 f^\mu\right) = 0\,.
}{eq:EW30}
The Bianchi identities establish a covariant conservation equation
\eq{
d_\mu \,j^\mu = 0
}{eq:EW31}
for the vector current
\eq{
j^\mu = r^{\mu\nu}f_\nu-\frac12 r f^\mu \pm \frac12 f^2 f^\mu\,.
}{eq:EW32}
This conservation is necessary and sufficient for \eqref{eq:EW30}. The 3-dimensional conservation \eqref{eq:EW31} of the current \eqref{eq:EW32} is recognized as the dimensionally reduced, 4-dimensional conservation
\eq{
D_A J^A = 0
}{eq:EW40}
of the Chern-Simons current
\eq{
J^A = \epsilon^{ABCD}\left(\Gamma^E{}_{BF} \,\partial_C \,\Gamma^F{}_{DE} 
+ \frac23 \Gamma^E{}_{BF}\,\Gamma^F{}_{CG}\,\Gamma^G{}_{DE} \right)
}{eq:EW39}
when $\pont$ vanishes. The structure of the current \eqref{eq:EW32} resembles the dimensionally reduced gravitational Chern-Simons term \cite{Guralnik:2003we}: 
it has a term cubic in $f$ and terms linear in $f$ which are coupled linearly to curvature.

We can now rephrase the constraint \eqref{eq:EW26} as the statement that the current \eqref{eq:EW32} must be covariantly conserved. A special case emerges if $f^\mu$ vanishes, i.e., $a_\mu$ is pure gauge. Then the current vector \eqref{eq:EW32} vanishes and \eqref{eq:EW31} holds trivially. This happens e.g.~for stationary spacetimes which are also static.

\section*{Acknowledgments}

We thank D.~Calderbank and R.~Ward for discussions. One of us (RJ) would like to thank M.~Eastwood for drawing attention to Einstein-Weyl spaces at a conference and the anonymous referee of \cite{Jackiw:2007ge} for helpful comments.

This work is supported in part by funds provided by the U.S. Department of Energy (D.O.E.) under the cooperative research agreement DEFG02-05ER41360.
DG has been supported by project GR-3157/1-1 of the German Research Foundation (DFG) and by the Marie Curie Fellowship MC-OIF 021421 of the European Commission under the Sixth EU Framework Programme for Research and Technological Development (FP6). 


\providecommand{\href}[2]{#2}\begingroup\raggedright\endgroup

\end{document}